\documentclass[a4paper,11pt,pdftex]{article}

\usepackage{here}
\usepackage{subfig}
\usepackage[pdftex]{graphicx}
\usepackage{color}
\usepackage{braket}
\usepackage{amssymb, amsmath}
\usepackage{feynmf}
\usepackage{bm}
\usepackage[samesize]{cancel}
\usepackage{url}
\usepackage{jcappub}

\usepackage{hyperref}
\usepackage[samesize]{cancel}
\usepackage{fancyhdr} 

\newcommand{\tk}{\tilde{\kappa}}
 
\newcommand{\be}{\begin{eqnarray}}
\newcommand{\ee}{\end{eqnarray}}

\newcommand{\beq}{\begin{equation}}
\newcommand{\eeq}{\end{equation}}
\newcommand{\beqa}{\begin{eqnarray}}
\newcommand{\eeqa}{\end{eqnarray}}

\newcommand{\lmk}{\left(}
\newcommand{\rmk}{\right)}
\newcommand{\lkk}{\left[}
\newcommand{\rkk}{\right]}
\newcommand{\lnk}{\left\{}
\newcommand{\rnk}{\right\}}

\newcommand{\la}{\langle}
\newcommand{\ra}{\rangle}

\title{\boldmath Renormalization in gravitational leptogenesis with pseudo-scalar-tensor coupling}

\author[a]{Kohei Kamada,}
\author[a,b]{Jun'ya Kume,}
\author[a]{and Yusuke Yamada}

\affiliation[a]{Research Center for the Early Universe (RESCEU), Graduate School of Science,\\ The University of Tokyo, Hongo 7-3-1
Bunkyo-ku, Tokyo 113-0033, Japan}
\affiliation[b]{Department of Physics, Graduate School of Science,
The University of Tokyo,\\ Hongo 7-3-1
Bunkyo-ku, Tokyo 113-0033, Japan}

\emailAdd{kohei.kamada@resceu.s.u-tokyo.ac.jp}
\emailAdd{kjun0107@resceu.s.u-tokyo.ac.jp}
\emailAdd{yamada@resceu.s.u-tokyo.ac.jp}

\subheader{{\rm RESCEU-12/20}}

\abstract{We consider the renormalization in the pseudo-scalar inflation models with the gravitational Chern-Simons term. In this model, 
lepton asymmetry is generated from the chiral gravitational waves produced due to the Chern-Simons term through the gravitational chiral anomaly. 
However, it is known that the naive estimate of the expectation value of the gravitational Chern-Pontryagin density as well as the 
resultant lepton number density depend on the UV-cutoff scale, 
which raises a question on their validity. 
In this paper, we propose a way to renormalize the expectation value of the Chern-Pontryagin density 
to remove the UV-cutoff dependence. 
We also discuss the renormalized lepton number density when we adopt the minimal subtraction scheme 
and the viability of the gravitational leptogenesis scenario.}

\begin{document}
\maketitle
\flushbottom

\section{Introduction}
The dynamical generation of the baryon asymmetry, so called baryogenesis, is one of the most plausible ways to explain the observed value of the baryon-to-entropy ratio of our universe, $n_B/s = (8.718\pm0.004)\times10^{-11}$~\cite{Aghanim:2018eyx}. Among various baryogenesis models, leptogenesis~\cite{Fukugita:1986hr} is considered as one of the most remarkable models. This is because the baryon asymmetry can be produced simply by adding right-handed Majorana neutrinos to the Standard Model of particle physics (SM), which simultaneously explains the non-zero neutrino mass measured in neutrino oscillation experiments through the type-I see-saw mechanism~\cite{Minkowski:1977sc,Yanagida:1980xy,GellMann:1980vs}. In the ``vanilla" leptogenesis scenario~\cite{Fukugita:1986hr}, the right-handed neutrinos are thermally produced during the reheating era and its decay results in the primordial lepton asymmetry, which is converted to the baryon asymmetry through the electroweak sphaleron process~\cite{Kuzmin:1985mm}. This scenario requires the reheating temperature to be $T_{\rm reh} \gtrsim 10^9$GeV, and the lightest right-handed neutrino mass to be $M_R \gtrsim 3 \times 10^9 {\rm GeV}$~\cite{Buchmuller:2004nz}. However, this bound can be relaxed by generating primordial lepton asymmetry with a different mechanism, such as the resonant leptogenesis~\cite{Pilaftsis:2003gt} or the leptogenesis
via active-sterile neutrino oscillation~\cite{Akhmedov:1998qx,Asaka:2005pn}.

In this paper, we focus on yet another leptogenesis scenario, so-called gravitational leptogenesis~\cite{Alexander:2004us} in which 
the absence of the right-handed neutrinos in the low-energy particle spectrum plays an essential role. 
Since the right-handed neutrinos do not exist in the SM, the lepton current does not conserved due to the gravitational chiral anomaly~\cite{AlvarezGaume:1983ig}.
Then if the chiral gravitational waves are produced so that the Chern-Pontryagin density $R{\tilde R}$ obtains a nonzero expectation value, 
the lepton asymmetry is induced through the anomaly equation. 
Chiral gravitational waves are generated 
in the models where pseudo-scalar inflaton $\phi$ couples to the Chern-Pontryagin density of the metric, $\phi R\tilde{R}$~\cite{Alexander:2004us, Lyth:2005jf, Fischler:2007tj,Kawai:2017kqt}, or the gauge fields, $\phi F\tilde{F}$\footnote{Baryogenesis mechanism that generates the asymmetry directly by the SM hypergauge fields (but not the chiral gravitational waves) through the SM chiral anomaly 
in the context of pseudo-scalar inflation has also been studied in Refs.~\cite{Bamba:2006km,Anber:2015yca,Jimenez:2017cdr,Domcke:2018eki,Domcke:2019mnd} as a different mechanism.}~\cite{Noorbala:2012fh, Maleknejad:2014wsa, Caldwell:2017chz, Papageorgiou:2017yup, Adshead:2018doq} for this scenario. 
These models might be tested by observing chirality of primordial gravitational waves~\cite{Romano:2016dpx, Smith:2016jqs}.

The coupling between a pseudo-scalar field and the Chern-Pontryagin density of a gauge field $\phi F\tilde{F}$ is often concerned~\cite{Turner:1987bw,Garretson:1992vt,Adshead:2012kp} especially in the framework of the natural inflation~\cite{Freese:1990rb}, 
since it is a common feature of the axion-like particles~\cite{Srednicki:1985xd}. 
On the other hand, 
the coupling to the gravitational Chern-Pontryagin density  $\phi R\tilde{R}$ is also allowed by symmetry, 
which we focus in the present work.  
Such a coupling can appear, {\it e.g.}, by the Green-Schwarz mechanism~\cite{Green:1984sg} in string theory. An interesting feature of the models with $\phi R\tilde{R}$ coupling is to generate chiral gravitational waves directly as vacuum fluctuation~\cite{Lue:1998mq,Choi:1999zy} rather than the gravitational waves sourced by helical gauge fields, which can be produced in the models with the Chern-Simons term $\phi F\tilde{F}$~\cite{Maleknejad:2012fw,Adshead:2013qp,Dimastrogiovanni:2012ew,Maleknejad:2016qjz,Dimastrogiovanni:2016fuu, Adshead:2016omu, Adshead:2017hnc}. Although the efficiency of the production of the lepton asymmetry in this model is constrained in order to avoid the appearance of ghost/strong coupling modes, the present authors have found that the model can explain the observed baryon asymmetry if the universe undergoes so called ``kination'' era~\cite{Spokoiny:1993kt,Joyce:1996cp} and sufficiently low reheating temperature is realized~\cite{Kamada:2019ewe}.

However, this model has a subtle issue in the evaluation of the net lepton number density. According to the previous works~\cite{Alexander:2004us, Lyth:2005jf}, a straightforward calculation tells that the expectation value of the generated lepton asymmetry as well as the gravitational Chern-Simons term depends on the UV cutoff of the momentum integral. 
The former has a physically reasonable UV cutoff, that is, the right-handed Majorana neutrino mass in the type-I see-saw mechanism, 
since gravitational chiral anomaly in the lepton current is canceled out at the high energy scale where the right-handed neutrinos enters the theory.  
On the other hand, the cut-off scale for the latter is identified to be the one for the effective theory such as the Planck scale. 
If such a UV-cutoff dependent expectation value of the gravitational Chern-Pontryagin density is physical and its UV cutoff scale 
is larger than the right-handed neutrino mass, 
the resultant cutoff-dependent lepton asymmetry may be justified. 
Nevertheless, one should note that this argument has an ambiguity since the UV divergent expectation value of the Chern-Pontryagin density 
depends on the regularization scheme. 
In other words, this cut-off dependence appears only when we use the cut-off regularization. 
Although it should depend on the detail of the UV completion of the model, 
one may think that such a UV cutoff dependent contribution should be removed by renormalization. 
The purpose of the present paper is to show if such a UV-dependent expectation value can be removed 
by appropriate renormalization scheme and how it affects the evaluation of the net lepton asymmetry.


We should emphasize that this issue was first considered in Ref.~\cite{Fischler:2007tj}, by using the analogy of the non-zero temperature quantum field theory to derive the effective action. 
From that point of view, it was argued that the resultant expectation value of the Chern-Pontryagin density does not have UV cutoff dependence 
after an appropriate renormalization. 
However, in the literature the renormalization procedure is not explicitly shown and it is not clear how the net lepton asymmetry is determined, 
which is supposed to depend on the renormalization conditions (or equivalently the renormalization scheme).
In this work, we clarify the structure of the UV divergences as well as the finite pieces in the expectation value of the Chern-Pontryagin density in this scenario.  The latter is identified to be the genuine physical one if we apply the minimal subtraction scheme with the cutoff regularization, which is consistent with the result of Ref.~\cite{Fischler:2007tj}. 
We follow the regularization procedure adopted in~\cite{Ballardini:2019rqh} where the authors considered the regularization of energy momentum tensor and helicity integral within the model containing $\phi F\tilde{F}$ coupling with the U(1) gauge fields in order to evaluate the backreaction of vector field production in the pseudo-scalar inflation~\cite{Turner:1987bw,Garretson:1992vt}. 
We identify the counter terms  that are needed for the minimal subtraction for the present scenario of pseudo-scalar inflation with the gravitational Chern-Simons term and evaluate the net baryon asymmetry when we adopt the scheme. 

The paper is organized as follows. In Sec.~\ref{review}, we review the gravitational leptogenesis scenario where pseudo-scalar inflaton couples to the gravitational Chern-Pontryagin density. Then we investigate the UV divergent structure of the expectation value of $R\tilde{R}$ in Sec.~\ref{reg} by adopting  the cutoff regularization as well as the adiabatic regularization~\cite{Zeldovich:1971mw,Parker:1974qw}. The detailed calculations of the contour integrals performed to find the UV divergences are shown in~\ref{appendix}. Then we consider renormalization of this term in Sec.~\ref{ren}, and discuss the implication to the lepton asymmetry generated in this model. Sec.~\ref{discuss} is devoted to summary and discussion.

\section{Review of the gravitational leptogenesis}\label{review}
\subsection{Linearized gravity action in the pseudo-scalar inflation}
Let us first review the standard evaluation of the generation of the chiral gravitational waves and lepton asymmetry 
following Ref.~\cite{Alexander:2004us}, which we have also adopted in our previous study Ref.~\cite{Kamada:2019ewe}, 
and clarify the issues in it. 
We consider the following model in which a pseudo-scalar field $\phi$ couples to the gravitational Chern-Simons term
\beq
\begin{split}
S &= S_{{\rm E}-{\rm H}} + S_{\rm gCS} + S_{\phi} + S_{\rm matter}\\
  &= \int d^4x\sqrt{-g}\left[\frac{M_{\rm Pl}^2}{2}R+\frac{M_{\rm Pl}^2}{4}f(\phi)R\tilde{R} + \mathcal{L}_{\phi} + \mathcal{L}_{\rm matter}\right],\label{action}
\end{split}
\eeq
which is the minimal setup for the gravitational leptogenesis. 
Here the gravitational Chern-Pontryagin density is given by
\beq
R\tilde{R} \equiv R_{\mu\nu\alpha\beta}\tilde{R}^{\mu\nu\alpha\beta} = \frac{1}{2}\frac{\epsilon^{\alpha\beta\gamma\delta}}{\sqrt{-g}}R_{\alpha\beta\rho\sigma}R_{\gamma\delta\mu\nu}g^{\mu\rho}g^{\nu\sigma}.
\eeq
In this paper, we adopt the metric convention $g_{\mu\nu} = (-,+,+,+)$ and $\epsilon^{\alpha \beta \gamma \delta}$
is the Levi-Civita tensor with $\epsilon^{0123}=1$. $M_\mathrm{Pl}\simeq2.43 \times 10^{18}$ GeV is the reduced Planck mass. 
We treat the pseudo-scalar inflaton $\phi$  as a homogeneous background field, namely, $\phi(\eta, {\bm x}) = \phi(\eta)$ where $\eta$ stands for the conformal time. 
We explicitly write the Lagrangian of the matter fields ${\cal L}_\mathrm{matter}$ that nontrivially interact with the gravity sector
through the gravitational chiral anomaly. 

On the perturbed Friedmann spacetime, 
\begin{equation}
ds^2 = a^2(\eta) [-d\eta^2+(\delta_{ij} + h_{ij}(\eta, {\bm x}))dx^i dx^j], 
\end{equation}
the gravity action up to the quadratic order in the tensor perturbation $h_{ij}$ is given by
\begin{equation}
\begin{split}
  {\cal S}_{\rm GW}^{(2)} = \frac{M_{\rm Pl}^2}{8}\int d^4x & \left[ a^2(\eta)\lnk(h^i_{\ j})'(h^j_{\ i})' - (\partial_kh^i_{\ j})(\partial^kh^j_{\ i})\rnk \right. \\ 
    & \left.- f'\epsilon^{ijk}\lnk(h^q_{\ i})'(\partial_jh_{kq})' - (\partial^rh^q_{\ i})\partial_j\partial_rh_{kq}\rnk \right], \label{S2}
\end{split}
\end{equation}
where the prime denotes the derivative with respect to the conformal time $\eta$. We take the transverse traceless (TT) gauge $h_{ii}=0$ and $\partial_i h_{ij}=0$.
It is convenient to move to the  Fourier space representation of $h_{ij}$ with the circular polarization tensors $p_{ij}^{L/R}({\bm k})$, 
\begin{equation}
h_{ij}(\eta,\textbf{x}) = \frac{1}{(2\pi)^{3/2}}\int d^3k\sum_{s = \mathrm{R,L}}p_{ij}^s(\bm{k})h_{\bm{k}}^s(\eta)e^{i\bm{k}\cdot \bm{x}}, \label{Fourier}
\end{equation}
where the circular polarization tensors satisfy the following equations, 
\begin{equation}
\begin{split}
  p_{ij}^\mathrm{R}({\bm k})p^{ij\mathrm{R}}({\bm k}) &= p^\mathrm{L}_{ij}({\bm k})p^{ij\mathrm{L}}({\bm k}) = 0,\\
  p_{ij}^\mathrm{R}({\bm k})p^{ij\mathrm{L}}({\bm k}) &= 2,\\
  k_p\epsilon^{mpj} p^A_{ij} ({\bm k}) &= -i\lambda^A_{\bm k} k \ p^{m \  A}_{\ \ i}({\bm k}) \ (\text{for} \ A=\mathrm{L, R}), 
\end{split} \label{cptr}
\end{equation}
with $\lambda^\mathrm{R}_{\bm k} = +1, \lambda^\mathrm{L}_{\bm k} = -1$. The polarization tensors satisfy
$p^A_{ij}({\bm k}) = p^A_{ij}(-{\bm k}) \ (A = \mathrm{L,R})$
so that the graviton satisfies the reality condition $h_{{\bm k}}^\mathrm{R} = (h_{-{\bm k}}^\mathrm{L})^*$.
Using these relations, the action (\ref{S2}) can be rewritten as
\begin{equation}
{\cal S}_{\rm GW}^{(2)} = \frac{M_{\rm Pl}^2}{4} \int d\eta d^3k \sum_{A = L,R} a^2(\eta)\left[1 - \lambda^A_{\bm k} k\frac{f'}{a^2(\eta)}\right](|(h^A_{{\bm k}})'|^2 - k^2|h^A_{{\bf k}}|^2).  \label{S2_f}
\end{equation}
Note that $\lambda_{\bm {k}}^A$ changes its sign as $\lambda^\mathrm{R}_{-\bm k} = -\lambda^\mathrm{R}_{\bm k} = -1, \lambda^\mathrm{L}_{-\bm k} = -\lambda^\mathrm{L}_{\bm k} = +1$, with respect to the reflection of the wave vector.
In order to clarify the issues we explore, here we assume the de Sitter expansion of the Universe, $a(\eta) = -1/(H\eta)$, 
with the Hubble parameter $H$ being constant, namely neglecting the slow-roll parameter corrections, and 
constant roll of inflaton, $f'/a =$ const. 
Note that here in the quadratic action for the metric perturbation, the inflaton field in $f(\phi)$ is taken as the classical background field but not the quantum field. 
Then the action is further simplified as
\begin{equation}
{\cal S}_{\rm GW}^{(2)} = \frac{1}{2} \int d\eta d^3k \sum_{A = L,R} z_A^2({\bm k})(|(h^A_{{\bm k}})'|^2 - k^2|h^A_{{\bf k}}|^2).  \label{S2_fz}
\end{equation}
where we have defined 
\begin{equation}
z_A^2(\eta, {\bm k}) = \frac{a^2M^2_{\rm Pl}}{2}\left(1 - \lambda^A_{\bm k} k\frac{f'}{a^2}\right) =  \frac{a^2M^2_{\rm Pl}}{2}\lmk1 - \lambda^A_{\bm k} k \frac{\Theta}{8} \eta\rmk,\quad \Theta \equiv -8H f'(\phi)/a = \mathrm{const}.
\end{equation}
Hereafter we take $\Theta>0$, without loss of generality. If we take $\Theta<0$, the same would apply  for the right-handed modes.
Since $z_A$ becomes negative for the left-handed polarization modes if $k>-8/(\Theta \eta)$, 
the model makes sense only for $k<-8/(\Theta \eta)$ as the kinetic term of higher momentum modes becomes ghost-like. 
In terms of the physical momentum $k_\mathrm{phys} \equiv k/a(\eta)$, the condition reads  
\begin{equation}
k_\mathrm{phys} < \frac{8 H}{\Theta}. 
\end{equation}
One may wonder  if the theory is catastrophic since it suffers from the ghost at high momenta. 
However, the theory is originally a non-renormalizable theory and has a cut-off scale above which the perturbative calculation breaks down. 
Indeed, the perturbative expansion up to quadratic order would be justified only for the physical momentum $k_\mathrm{phys}<M_\mathrm{Pl}$. 
For larger $k_\mathrm{phys}>M_\mathrm{Pl}$, nonlinear interactions of the metric perturbation is no longer negligible
and the system is strongly coupled so that we cannot predict anything on these scales. 
The Chern-Simons term also gives a strongly-coupled scale, but it depends on the detail of the function of $f(\phi)$. 
Thus conservatively we can regard that the apparent ghost-like kinetic term at the scale above the UV-cutoff scale, $\Lambda = M_\mathrm{Pl}$, 
is not physical and not catastrophic. 
This argument just gives an upper bound of $\Theta$ as $\Theta< 8 H /M_\mathrm{Pl} \lesssim 10^{-5}$~\cite{Alexander:2004wk}  
where we have used the observational
upper bound of the Hubble parameter during inflation, $H \lesssim 10^{13}$ GeV~\cite{Aghanim:2018eyx}. Hereafter we consider the linear function $f(\phi)$
\beq
f(\phi) = \frac{\mathcal{N}}{16\pi^2M_{\rm Pl}^2}\frac{\phi}{M_{\rm Pl}},
\eeq
which respects the shift symmetry of the pseudo-scalar $\phi$ in order to make the discussion clear and concrete. 

Let us now discuss the mode equations of the gravitational waves. Defining canonical variables as
\begin{equation}
\mu_{\bm k}^A \equiv z_A({\bm k}) h_{\bm k}, 
\end{equation}
the action is further rewritten as
\begin{equation}
{\cal S}_{\rm GW}^{(2)} = \frac{1}{2} \int d\eta d^3k \sum_{A = L,R} \left(|(\mu^A_{{\bm k}})'|^2- \left(k^2-\frac{z_A''({\bm k})}{z_A({\bm k})}\right)|\mu^A_{{\bf k}}|^2\right),  \label{S2_fm}
\end{equation}
and the mode equation reads
\begin{equation}
 (\mu_{\bm{k}}^A)'' +\left(k^2 - \frac{z_A''({\bm k})}{z_A({\bm k})}\right)  \mu_{\bm{k}}^A = 0,\label{modeeq0} 
\end{equation}
or equivalently
\begin{equation}
 (\mu_{\bm{k}}^A)'' +\left(k^2  - \frac{2}{\eta^2}-\frac{\lambda^A_{\bm k} k \Theta/8}{(1-\lambda^A_{\bm k} k \Theta \eta/8)\eta}+\frac{k^2 \Theta^2/256}{(1-\lambda^A_{\bm k} k \Theta \eta/8)^2}\right)\mu_{\bm k}^A =0,  \label{modeeq}
\end{equation}
with $\Theta \lesssim 8 H / M_\mathrm{Pl}$ for the validity of the model.

We quantize the graviton by decomposing $\mu_{\bm k}^A$ as the quantum operators so that
\begin{equation}
\begin{split}
  \hat{\mu}_{{\bm k}}^\mathrm{R} (\eta)&= u_{{\bm k}}^\mathrm{R} (\eta)\hat{a}_{{\bm k}} + (u_{-{\bm k}}^\mathrm{L} (\eta))^*  \hat{b}^{\dagger}_{-{\bm k}},\\
  \hat{\mu}_{{\bm k}}^\mathrm{L}  (\eta) &= u_{{\bm k}}^\mathrm{L}  (\eta) \hat{b}_{{\bm k}} + (u_{-{\bm k}}^\mathrm{R} (\eta))^*\hat{a}^{\dagger}_{-{\bm k}},\label{mode}
\end{split}
\end{equation}
where $\hat{a}_{{\bm k}}$ and $\hat{a}^{\dagger}_{{\bm k}}$ denote the annihilation and creation operator of the right-polarized mode with momentum ${\bm k}$, and $\hat{b}_{{\bm k}}$ and $\hat{b}^{\dagger}_{{\bm k}}$ denote those of the left-polarized mode, respectively. These operators satisfy commutation relations, $[\hat{a}_{{\bm k}}, \hat{a}^{\dagger}_{{\bm k'}} ] = [\hat{b}_{{\bm k}}, \hat{b}^{\dagger}_{{\bm k'}}] = \delta({\bm k} - {\bm k'})$.
Here we have used the hats to denote that they are quantum operators. 
Then we define the vacuum state $ |0\rangle$ that satisfies 
\begin{equation}
\hat{a}_{{\bm k}} |0\rangle  = \hat{b}_{{\bm k}} |0\rangle =0. 
\end{equation}

We would like to take the mode function in the Bunch-Davies-like vacuum that has the asymptotic form 
\begin{equation}
u_{\bm k}^A \simeq (\omega_k)^{-1/2} \exp\left[-i \int^\eta \omega_k(\eta') d\eta'\right],\quad \omega_k \equiv \sqrt{k^2+\frac{z_A''({\bm k})}{z_A({\bm k})}}, 
\end{equation}
at $k\eta \rightarrow - \infty$. 
However, the exact mode equation Eq.~\eqref{modeeq} is singular at $k\eta = - 8/\Theta$
and we cannot take the $k\eta \rightarrow -\infty$ limit. 
Then we require that the mode function consists only of the positive frequency mode at a time $\eta_\mathrm{i}$ when 
the mode is deep inside the horizon but $z_A$ is almost 1, $8/\Theta \gg - k\eta_\mathrm{i} \gg 1$~\cite{Alexander:2004wk}. 
By solving the equation of motion at when $z_A\simeq 1$, 
\begin{equation}
(u_{\bm k}^A)'' + \left(k^2 - \frac{2}{\eta^2}-\frac{\lambda^A_{\bm k} k \Theta}{8\eta}-\frac{3k^2 \Theta^2}{256}\right)u_{\bm k}^A =0, \label{approxeom}
\end{equation}
we obtain the positive frequency mode function as 
\begin{equation}
u_{\bm k}^A(\eta) = \frac{1}{\sqrt{k} }e^{i k \eta_\mathrm{i}} \exp\left[-\frac{\pi\lambda^A_{\bm k}\Theta}{32}\right] W_{\kappa, 3/2}\left(i \sqrt{4-\frac{3\Theta^2}{64}}k\eta \right), \quad \kappa \equiv \frac{i\lambda^A_{\bm k}}{\sqrt{256/\Theta^2-3}},  \label{whitsol}
\end{equation}
where  $W_{\kappa,\mu}(z)$ is the Whittaker function. 
The asymptotic form of this solution is 
\begin{equation}
u_\mathrm{k}^A(\eta)  = \frac{1}{\sqrt{k}}  \exp\left[-\frac{\lambda^A_{\bm{k}} \pi \Theta}{32}\right]  \exp\left[-ik \left( \sqrt{1-\frac{3\Theta^2}{256}}\eta -\eta_\mathrm{i}\right)\right], 
\end{equation}
for $-k \eta_\mathrm{i} >  - k\eta \gg 1$, where we have used the asymptotic form of the Whittaker function at $z \rightarrow \infty$, 
\begin{equation}
W_{\kappa, \mu}(z) \simeq e^{-z/2} z^\kappa.
\end{equation}
Note that in the vanishing $\Theta$ limit, it is a simple positive frequency mode, 
\begin{equation}
  u_\mathrm{k}^A(\eta) = \frac{1}{\sqrt{k}} \exp[-ik(\eta-\eta_\mathrm{i})]. 
\end{equation} 
With the nonzero $\Theta$ the mode functions for the left-handed and right-handed polarization mode are different, 
which suggests the generation of chiral gravitational waves. 
In the following, we will use the linearized gravity action and mode functions in evaluating the expectation value of 
the Chern-Pontryagin density $R\tilde{R}$, which is eventually converted to the lepton number.

\subsection{Lepton number production during inflation}\label{leptogenesis}
As pointed out in~\cite{Alexander:2004us}, lepton asymmetry can be produced during inflation through the gravitational anomaly~\cite{AlvarezGaume:1983ig} in the total lepton number current $J_{\rm L}^{\mu}$ via the relation
\beq
\nabla_{\mu}J_L^{\mu} = \frac{N_{\rm R-L}}{24(4\pi)^2}R\tilde{R},\label{G_anomaly}
\eeq
where $N_{\rm R-L}$ is the difference in the number of species of right- and left-handed leptons. If we assume that three heavy right-handed Majorana neutrinos exist with a common mass scale $M_R=\mu$, then $N_{\rm R-L} = -3$ below the energy scale $\mu$. Thus $\mu$ can be identified as the UV cutoff of the integration. Taking expectation value  and integrating this equation over the inflationary era, we can evaluate the total lepton asymmetry at the end of inflation.

The evaluation is performed as follows. The gravitational Chern-Pontryagin density $R\tilde{R}$ can be written as
\beq
a^4R\tilde{R} = \partial_{\mu}(a^4K^{\mu}) = \partial_{\eta}\lkk\frac{1}{2}\epsilon^{ijk}(-\partial_lh_{jm}\partial_m\partial_ih_{kl} + \partial_lh_{jm}\partial_l\partial_ih_{km} - h'_{jl}\partial_ih'_{lk})\rkk + \cdots,\label{RR_dual}
\eeq
where the ellipses denote the spatial total derivative terms, which vanish by taking spatial average. The first term also vanishes in the TT gauge. 
The expectation values of the second and the third terms can be expressed as follows,
\beq
\langle\epsilon^{ijk}\partial_m\hat{h}_{jl}\partial_i\partial_m\hat{h}_{lk}\rangle
= 2\int\frac{d^3k}{(2\pi)^3}k^3\left[-\frac{u_{\bm k}^R}{z_R({\bf k})}\frac{(u_{\bm k}^R)^*}{z_L(-{\bm k})} + \frac{u_{\bm k}^L}{z_L({\bf k})}\frac{(u_{\bm k}^L)^*}{z_R(-{\bm k})}\right],
\eeq

\beq
\langle\epsilon^{ijk}\hat{h}^{\prime}_{jl}\partial_i\hat{h}^{\prime}_{lk}\rangle
= 2\int\frac{d^3k}{(2\pi)^3}k\left[-\left(\frac{u_{\bm k}^R}{z_R({\bf k})}\right)^{\prime}\left(\frac{(u_{\bm k}^R)^*}{z_L(-{\bm k})}\right)^{\prime} + \left(\frac{u_{\bm k}^L}{z_L({\bf k})}\right)^{\prime}\left(\frac{(u_{\bm k}^L)^*}{z_R(-{\bm k})}\right)^{\prime}\right].
\eeq
Combining these expressions, we obtain the expectation value of $R\tilde{R}$ as
\beq
\langle{R\tilde{R}}\rangle = \frac{1}{a^4}\partial_{\eta}\left[\int_{k<\Lambda a}\frac{d^3k}{(2\pi)^3}\sum_{A = R,L}\frac{k\lambda^A_{\bm{k}}}{z^2_A({\bm{k}})}\left\{|u^{A\prime}_{\bm k}|^2 - \left(k^2 - \left(\frac{z^{\prime}_A({\bm{k}})}{z_A({\bm{k}})}\right)^2\right)|u^{A}_{\bm k}|^2 - \frac{z^{\prime}_A({\bm{k}})}{z_A({\bm{k}})}(u^{A}_{\bm k}u^{A*\prime}_{\bm k}+u^{A\prime}_{\bm k}u^{A*}_{\bm k})\right\}\right],  \label{RR_VEV}
\eeq
where we have adopted the cutoff regularization where we have the upper bound of the $k$ integration as $k<\Lambda a$. 
Note that we can omit the ${\bm k}$ dependence of $\lambda^A$ and $z_A$ with $\lambda^R = +1, \lambda^L = -1$, since we do not consider the reflection of the wave vector in the following.  
With Eq.~\eqref{RR_VEV}, let us perform the integration of Eq.~\eqref{G_anomaly} and take the spatial average. We obtain the lepton number produced during inflation as
\beq
\langle n_{L}(\eta_f)\rangle = \frac{N_{\rm R-L}}{384\pi^2a_f^3}\int_{k<\mu a}\frac{d^3k}{(2\pi)^3}\sum_{A = R,L}\frac{k\lambda^A}{z^2_A}\left\{|u^{A\prime}_{\bm k}|^2 - \left(k^2 - \left(\frac{z^{\prime}_A}{z_A}\right)^2\right)|u^{A}_{\bm k}|^2 - \frac{z^{\prime}_A}{z_A}(u^{A}_{\bm k}u^{A*\prime}_{\bm k}+u^{A\prime}_{\bm k}u^{A*}_{\bm k})\right\}.
\eeq
Here the upper bound of the $k$ integration is replaced by $\mu a$ as discussed in the above. 
Substituting the mode function~\eqref{whitsol} and picking the leading divergence, the lepton asymmetry is evaluated as
\beq
\langle n_{L}(\eta_f)\rangle = -\frac{1}{2048\pi^4}\lmk\frac{H}{M_{\rm Pl}}\rmk^2\Theta H^3\lmk\frac{\mu}{H}\rmk^4, \label{lepton}
\eeq
at the leading order in $\Theta$. 
The lepton asymmetry is eventually converted into the baryon asymmetry through the electroweak sphalerons~\cite{Kuzmin:1985mm} as $n_B = -28/79 n_L$ 
before the electroweak symmetry breaking~\cite{Harvey:1990qw}. (Particle contents of the SM is assumed.)
If the effective equation of state during reheating is $p = \rho$, 
which can be realized {\it e.g.}, in the kinetically driven inflation model~\cite{ArmendarizPicon:1999rj,Kobayashi:2010cm} with gravitational reheating~\cite{Ford:1986sy,Kunimitsu:2012xx,Figueroa:2016dsc,Nakama:2018gll,Dimopoulos:2018wfg,Hashiba:2018iff}, 
the resultant baryon asymmetry becomes comparable to the observed value $n_B/s \sim 8.7\times10^{-11}$ as~\cite{Kamada:2019ewe}
\beq
\frac{n_B}{s} = 9.7 \times 10^{-11} \lmk\frac{\epsilon_{\rm w.o}}{1}\rmk\lmk\frac{g_*}{100}\rmk^{-1/2}\lmk\frac{\mu}{10^{16}\mathrm{GeV}}\rmk^4\lmk\frac{\Theta}{10^{-5}}\rmk\lmk\frac{T_{\rm reh}}{10^7\mathrm{GeV}}\rmk^{-1}. \label{baryon_1}
\eeq
Here $g_*$ is the effective number of relativistic degrees of freedom at the reheating, which is ${\cal O}(100)$ unless we do not consider extremely large 
number of particle species in the physics beyond the SM above the electroweak scale,  
and $\epsilon_{\rm w.o}\lesssim 1$ is a washout factor taking into account the lepton number violating process due to the Majorana mass term, which turns out to be irrelevant for this case~\cite{Adshead:2017znw}. 
Note that the right-handed neutrino mass has an upper bound as $\mu <10^{16}$ GeV from
the requirement of the perturbativity of the neutrino Yukawa interactions~\cite{Abazajian:2012ys}.
Reheating temperature cannot be much smaller than $10^{7}$ GeV in this case since otherwise the abundance of the gravitationally produced 
gravitons becomes too much and inconsistent with the constraint from the successful Big Bang Nucleosynthesis (BBN). 

However, this evaluation of the lepton number production~\eqref{lepton} has subtleties as discussed in the introduction. We can easily see 
that the resultant lepton number, as well as the Chern-Pontryagin density, 
is a divergent quantity, although, on the former, the cutoff scale $\mu$ can be originated from a well-motivated physical scale. 
It is apparent that this evaluation depends on the regularization method since Eq.~\eqref{lepton} explicitly relies on the cutoff regularization. 
As is well known, in order to evaluate a divergent quantity physically in quantum theories, 
one needs to renormalize it after regularization. 
Thus we also need to perform the renormalization in this case, as also argued in Ref.~\cite{Fischler:2007tj}.  
In the next section, we will explicitly show the UV divergent structure as well as the finite part in the evaluation of the expectation value 
of  the Chern-Pontryagin density 
$\langle R\tilde{R}\rangle$ so that it is clearer how we can renormalize it.  

\section{UV divergent structure and regularizations}\label{reg}
\subsection{UV divergence of the expectation value of gravitational Chern-Pontryagin density}\label{ana}
We now evaluate the expectation value of the Chern-Pontryagin density $R\tilde{R}$ with the mode function~\eqref{whitsol} by using the cutoff regularization
along the line performed in Ref.~\cite{Ballardini:2019rqh} to determine the subleading and finite contributions. 
First, we expand Eq.~\eqref{RR_VEV} up to leading order of $\Theta$ as
\beq
\begin{split}
  \langle{R\tilde{R}}\rangle = &\frac{1}{a^4}\partial_{\eta}\left[\eta^{-4}\sum_A\frac{2\lambda^A}{M^2_{\rm Pl}a^2}\frac{1}{2\pi^2}\int_{x<\frac{\Lambda}{H}}dx x^3\left(1 - \lambda^A\frac{\Theta}{8}x+\mathcal{O}\lmk\lmk\frac{\Theta}{8}x\rmk^2\rmk\right)\times\right.\\
&\left.\left\{|u^{A\prime}_{\bm k}|^2 - \left(x^2 +\lambda^A\frac{\Theta}{8}x - 1\right)\eta^{-2}|u^{A}_{\bm k}|^2 + \left(1 - \lambda^A\frac{\Theta}{16}x\right)\eta^{-1}(u^{A}_{\bm k}u^{A*\prime}_{\bm k}+u^{A\prime}_{\bm k}u^{A*}_{\bm k})\right\}\right],\label{RR}
\end{split}
\eeq
where $x \equiv -k\eta$. 
Here the upper bound of the $k$ integration is taken as $- \Lambda a \eta= \Lambda/H$. 
Note that the previous calculation~\eqref{lepton} has taken into account up to the leading divergence in $\Lambda$. 
In order to find all the UV divergent part as well as the finite contribution, 
it is useful to take the Mellin-Barnes representation of the Whittaker function,
\beq
W_{\kappa, \mu}(z) = \frac{1}{2\pi i}e^{-z/2}\int_{\mathcal C_s}ds z^s\frac{\Gamma(s-\kappa)\Gamma(-s-\mu+1/2)\Gamma(-s+\mu+1/2)}{\Gamma(-\kappa-\mu+1/2)\Gamma(\mu-\kappa+1/2)},
\eeq
where the contour ${\mathcal C_s}$ runs from $-i\infty$ to $i\infty$ and is taken to separate the poles of $\Gamma(-s-\mu+1/2)$ and $\Gamma(-s+\mu+1/2)$ from those of $\Gamma(s-\kappa)$. 
For convenience, we define the products of Gamma function as
\beq
f(s, t, \kappa) \equiv \Gamma(s - i\kappa)\Gamma(-s-1)\Gamma(-s+2)\Gamma(t + i\kappa)\Gamma(-t -1)\Gamma(-t + 2),
\eeq
and the following quantity
\beq
\tilde{\kappa} \equiv 1/\sqrt{256/\Theta^2 -3} \simeq \frac{\Theta}{16}+\mathcal{O}(\Theta^3),
\eeq
\beq
\alpha \equiv \sqrt{4 - \frac{3}{64}\Theta^2} \simeq 2+\mathcal{O}(\Theta^2).
\eeq
The products of mode functions can be expressed as
\begin{align}
|u_{\bm k}^A|^2 = &\frac{1}{k}e^{-\pi\lambda^A\tilde{\kappa}}\frac{\sinh^2(\pi\tilde{\kappa})}{\pi^2}\int_{\mathcal C_s}\int_{\mathcal C_t}\frac{ds}{2\pi i}\frac{dt}{2\pi i}(i\alpha k\eta)^s(-i\alpha k\eta)^tf(s, t, \lambda^A\tilde{\kappa}),\\
|u_{\bm k}^{A\prime}|^2 =& \alpha^2ke^{-\pi\lambda^A\tilde{\kappa}}\frac{\sinh^2(\pi\tilde{\kappa})}{\pi^2}\int_{\mathcal C_s}\int_{\mathcal C_t}\frac{ds}{2\pi i}\frac{dt}{2\pi i}(i\alpha k\eta)^{s-1}(-i\alpha k\eta)^{t-1}\nonumber\\
&\qquad\times\left(s - \frac{i\alpha k}{2}\eta\right)\left(t + \frac{i\alpha k}{2}\eta\right)f(s, t, \lambda^A\tilde{\kappa}),\\
u_{\bm k}^Au_{\bm k}^{A\prime *} =& -i\alpha e^{-\pi\lambda^A\tilde{\kappa}}\frac{\sinh^2(\pi\tilde{\kappa})}{\pi^2}\int_{\mathcal C_s}\int_{\mathcal C_t}\frac{ds}{2\pi i}\frac{dt}{2\pi i}(i\alpha k\eta)^{s}(-i\alpha k\eta)^{t-1}\left(t + \frac{i\alpha k}{2}\eta\right)f(s, t, \lambda^A\tilde{\kappa}),\\
u_{\bm k}^{A*}u_{\bm k}^{A\prime} =& i\alpha e^{-\pi\lambda^A\tilde{\kappa}}\frac{\sinh^2(\pi\tilde{\kappa})}{\pi^2}\int_{\mathcal C_s}\int_{\mathcal C_t}\frac{ds}{2\pi i}\frac{dt}{2\pi i}(i\alpha k\eta)^{s-1}(-i\alpha k\eta)^{t}\left(s - \frac{i\alpha k}{2}\eta\right)f(s, t, \lambda^A\tilde{\kappa}),
\end{align}
where we have used
\beq
\frac{1}{\Gamma(-i\lambda^A\tilde{\kappa}-1)\Gamma(i\lambda^A\tilde{\kappa}+2)} = i\frac{\sinh(\pi\lambda^A\tilde{\kappa})}{\pi}.
\eeq
Combining them all,  
by performing the $x$ integration, we obtain the expectation value of the Chern-Pontryagin density $R\tilde{R}$ in terms of  the contour integral as
\beq
\begin{split}
\langle &R\tilde{R}\rangle = 3H^4\frac{\sinh^2(\pi\tilde{\kappa})}{\pi^4}\frac{H^2}{M^2_{\rm Pl}}\int_{\mathcal C_s}\int_{\mathcal C_t}\frac{ds}{(2\pi i)}\frac{dt}{(2\pi i)}f(s,t,\tilde{\kappa})\alpha^{s+t}\times\\
  &\left\{\left(e^{-\pi\tilde{\kappa}-i\frac{\pi}{2}(s - t)} - e^{\pi\tilde{\kappa}+i\frac{\pi}{2}(s - t)}\right)\left[\frac{(s+1)(t+1)}{s+t+3}x^{s+t+3}+\frac{\Theta}{8}i\frac{s-t}{s+t+5}x^{s+t+5}\right]^{\frac{\Lambda}{H}}_{\frac{c}{H}} \right.\\
  &-\left.\left(e^{-\pi\tilde{\kappa}-i\frac{\pi}{2}(s - t)} + e^{\pi\tilde{\kappa}+i\frac{\pi}{2}(s - t)}\right)\left[i\frac{s-t}{s+t+4}x^{s+t+4}+\frac{\Theta}{16}\frac{2(s+1)(t+1)+s+t+2}{s+t+4}x^{s+t+4}\right]^{\frac{\Lambda}{H}}_{\frac{c}{H}}\right\}, \label{integrand}
\end{split}
\eeq
where we have used the following relation
\beq
f(s,t,\lambda^R\tilde{\kappa}) = f(t, s, \lambda^L\tilde{\kappa}),
\eeq
and introduced the IR regulator $c$ together with the UV cutoff $\Lambda$. 
Note that the IR divergence does not appear in above calculation and we can safely take $c\to0$. One can explicitly check it using \eqref{integrand}.

By performing the contour integral with respect to $s$ and $t$, finally we obtain the regularized expectation value  of the Chern-Pontryagin density  $R\tilde{R}$  as
\beq
  \langle R\tilde{R}\rangle = 3H^4\frac{H^2}{\pi^2M_{\rm Pl}^2}\tk\lnk\lmk\frac{\Lambda}{H}\rmk^4+3\lmk\frac{\Lambda}{H}\rmk^2-10\log\lmk\frac{2\Lambda}{H}\rmk-10\gamma+\frac{39}{2}\rnk+\mathcal{O}(\tk^2).\label{RRlambda}
\eeq
The detail of the calculation is shown in appendix~\ref{appendix} for readers who are interested in. As a result of the calculation, we have explicitly shown the UV cutoff dependence including the quadratic and the logarithmic divergences, which was not explicitly shown in the previous studies. 

In addition, the result of the calculation includes the finite part of the expectation value  of the Chern-Pontryagin density $R\tilde{R}$. This is relevant for the evaluation of the renormalized lepton number density if we adopt a minimal subtraction scheme. 
We note that the finite part we have shown would depend on the choice of the vacuum, and therefore, one can say that the finite part is the prediction of the vacuum state we have chosen. 

On the other hand, it is known that if the vacuum has the Hadamard property, the UV divergence structure is the same. (See e.g. Ref.~\cite{Hack:2015zsu}.) In our case, the mode function is ill-defined at high energy limit, and in this sense, we may say that the vacuum is quasi-Hadamard state. Nevertheless, as we will show in the next subsection, the adiabatic vacuum, which we may also call a quasi-Hadamard state, can reproduce precisely the same UV divergences we show here. Thus we conclude that the UV divergent part calculated here is a general consequence of a reasonable choice of the vacuum. 

We also note that the UV divergences we have specified  is derived under a specific background (constant $H$ and $\Theta$). This procedure is not covariant under diffeomorphism, and covariant procedure of regularization such as Schwinger-DeWitt method~\cite{DeWitt:1975ys} (see also Ref.~\cite{Birrell:1982ix}) would be necessary if we would like to recover the general covariance.\footnote{However, for general FRW background, the adiabatic regularization, which seems not covariant at first glance, is known to be equivalent to the Schwinger-DeWitt point-splitting method, up to the adiabatic order necessary for renormalization. See e.g. appendix of~\cite{Anderson:1987yt}. In our case, since there is a background scalar field, the relation to the covariant expression is not clear.} Nevertheless, the regularization under this specific background is enough for our phenomenological purpose.

\subsection{Comparison to the adiabatic regularization}\label{ad_reg}
In this section, we perform the adiabatic regularization~\cite{Zeldovich:1971mw,Parker:1974qw} of the expectation value of the gravitational Chern-Pontryagin density $\langle R\tilde{R}\rangle$ as an independent method to obtain the UV divergent part.\footnote{See also \cite{Parker:2009uva} for a review of adiabatic regularization.} 
In the standard procedure of the adiabatic regularization, we first construct the ``adiabatic'' mode function $u^A_{\rm ad}(k,\eta)$ with the form of 
\beq
u^A_{\rm ad}(k,\eta) = \frac{1}{\sqrt{\Omega_A(k,\eta)}}\exp\lnk  i\int^{\eta}d\eta'\Omega_A(k,\eta')\rnk, \label{ad_mode}
\eeq
where $\Omega_A (k,\eta)$ is the adiabatic frequency which is determined so that $u^A_{\rm ad}(k,\eta)$ is the formal solution of Eq.~\eqref{modeeq0}. 
By substituting this mode function into the mode equation~\eqref{modeeq0}, we obtain the following relation,
\beq
\Omega_A^2 = \bar{\Omega}_A^2 + \frac{3}{4}\lmk\frac{\Omega'_A}{\Omega_A}\rmk^2 - \frac{1}{2}\frac{\Omega_A^{\prime\prime}}{\Omega_A},\label{ad_freq} 
\eeq
where
\beq
\bar{\Omega}_A^2(k,\eta) \equiv k^2 - \frac{z_A^{\prime\prime}(k,\eta)}{z_A(k,\eta)}.
\eeq
With this formal solution, the expectation value of the Chern-Pontryagin density $R\tilde{R}$ with respect to this adiabatic vacuum can be calculated 
in terms of the adiabatic frequency $\Omega_A$ as
\beq
  \langle{R\tilde{R}}\rangle_{\rm ad} = \frac{1}{a^4}\partial_{\eta}\lkk\int_{k<\Lambda a}\frac{d^3k}{(2\pi)^3}\sum_{A = R,L}\frac{\lambda^Ak}{\Omega_Az^2_A}\lnk \Omega_A^2 + \frac{1}{4}\lmk\frac{\Omega_A^{\prime}}{\Omega_A}\rmk^2 - \lmk k^2 - \left(\frac{z^{\prime}_A}{z_A}\right)^2\rmk + \frac{z^{\prime}_A}{z_A}\lmk\frac{\Omega_A^{\prime}}{\Omega_A}\rmk\rnk\rkk.
  \label{RRad}
\eeq

Now our task is to solve Eq.~\eqref{ad_freq}.  
In the adiabatic limit, namely in the limit the variation of the background whose effect appears in $z_A$ is sufficiently slow, 
we can solve it iteratively by introducing an adiabatic parameter $\epsilon$. 
Note that since the UV modes are hardly affected by the spacetime curvature, this solution~\eqref{ad_mode} in the slowly varying background should correctly reproduce the UV divergences.
The adiabatic expansion is done by assigning a power of the adiabatic parameter $\epsilon$  to each of the derivative with respect to $\eta$
and by determining the $2n$-th order adiabatic mode function recursively with $n$ iteration in Eq.~\eqref{ad_freq}. 
Namely, we write the adiabatically expanded frequency $\Omega_A$ as
\beq
\Omega_A^2  =   (\Omega_A^{(2n)})^2 + {\cal O}(\epsilon^{2(n+1)}), 
\eeq
with the following recurrence relation, 
\beq
\begin{split}
  (\Omega_A^{(0)})^2 &= k^2 - \epsilon^2\frac{z_A^{\prime\prime}}{z_A},\\
  (\Omega_A^{(2n)})^2 &= (\Omega_A^{(0)})^2 + \frac{3}{4}\epsilon^2\lmk\frac{\Omega^{(2n-2)\prime}_A}{\Omega^{(2n-2)}_A}\rmk^2 - \frac{1}{2}\epsilon^2\frac{\Omega_A^{(2n-2)\prime\prime}}{\Omega_A^{(2n-2)\prime}}\ \ \ (n \geq 1).
\end{split} 
\eeq
Note that $z_A$ also contains derivative with respect to $\eta$ so that we shall rewrite it with the adiabatic parameter $\epsilon$ as 
\beq
z_A^2 = \frac{a^2M_{\rm Pl}^2}{2}\lmk1 -\lambda^Ak\epsilon\frac{f^{\prime}(\phi)}{a^2}\rmk = \frac{a^2M_{\rm Pl}^2}{2}\lmk1 -\lambda^Ak\epsilon\frac{\Theta}{8}\eta\rmk.
\eeq

Expanding up to 8-th adiabatic order, we obtain all the UV divergent part of the expectation value of the Chern-Pontryagin density $R{\tilde R}$ as
\beq
\begin{split}
  \langle{R\tilde{R}}\rangle_{\rm ad} &= \frac{1}{a^4}\epsilon\partial_{\eta}\lkk a^{-3}\frac{H^5}{\pi^2M_{\rm Pl}^2}\tk\lnk\lmk\frac{\Lambda}{H}\rmk^4\epsilon^3+3\lmk\frac{\Lambda}{H}\rmk^2\epsilon^5-10\log\lmk\frac{2\Lambda}{H}\rmk\epsilon^7+10\log\lmk\frac{2c}{H}\rmk\epsilon^7+\mathcal{O}(\epsilon^9)\rnk\rkk\\
  &=3H^4\frac{H^2}{\pi^2M_{\rm Pl}^2}\tk\lnk\lmk\frac{\Lambda}{H}\rmk^4\epsilon^4+3\lmk\frac{\Lambda}{H}\rmk^2\epsilon^6-10\log\lmk\frac{2\Lambda}{H}\rmk\epsilon^8-10\log\lmk\frac{2c}{H}\rmk\epsilon^8\rnk,
\end{split} \label{RRad2}
\eeq
where we have introduced an IR regulator $c$ or a lower cutoff of the integration. 
We can see that  Eq.~\eqref{RRad2} contains the same divergences as in Eq.~\eqref{RRlambda} after taking $\epsilon\to1$. This result confirms our result~\eqref{RRlambda} derived with a different method as well as a different but similar vacuum in the previous section.

Compared to the previous calculation, however, one cannot determine the finite part of $R\tilde{R}$ in this calculation. Besides, the expectation  value of $R\tilde{R}$ depends on the IR regulator $c$. 
Since the adiabatic mode function is constructed to reproduce the generic feature of the UV modes, its validity breaks down for the IR region and 
it is reasonable that a logarithmic IR divergences appear. 
In general, $c$ can be taken as the scale where the adiabatic approximation breaks down, {\it e.g.} horizon scale during inflation~\cite{Durrer:2009ii, Marozzi:2011da}. However, as we have shown in Sec.~\ref{ana}, the local gauge invariant quantity $R\tilde{R}$ with an appropriate vacuum choice does not exhibit such (unphysical) IR divergence. 
Thus we here use the results with the adiabatic regularization just to show the generality of the UV divergence 
and will not explore the IR divergence as well as the finite part in depth. 
See Ref.~\cite{Faizal:2011iv} and references therein for discussion on the absence of the IR divergence of physical graviton propagator.

\section{Renormalization and its implication to leptogenesis}\label{ren}
\subsection{Counter terms}
In the previous section, we have explicitly shown the UV cutoff dependence of the expectation value of the Chern-Pontryagin density $R\tilde{R}$, which is independent of the choice of vacuum states with (quasi-)Hadamard property.  In this section, we propose the way to renormalize it by removing the UV cutoff dependent part. 
Let us identify the counter terms as follows. 
The expectation value of the Chern-Pontryagin density $R{\tilde R}$ physically appears in the effective equation of motion for $\phi$. 
Since the classical equation of motion for $\phi$ with a canonical kinetic term and a potential $V(\phi)$ is given by
\beq
  \Box \phi - V_{,\phi} = \frac{\mathcal{N}}{64\pi^2M_{\rm Pl}}R\tilde{R},\label{eom}
\eeq
after renormalization we require that the effective equation of motion becomes
 \beq
  \Box \phi_\mathrm{cl} - V_{,\phi}^\mathrm{eff} = \frac{\mathcal{N}}{64\pi^2M_{\rm Pl}}\langle R\tilde{R}\rangle_\mathrm{ren},\label{efeom}
\eeq
where $\phi_\mathrm{cl}$ is the classical field, $V^\mathrm{eff}$ is the effective potential,\footnote{Such effective potential would appear from interactions of inflaton and also contain UV divergences, which needs to be regularized. We will not discuss the regularization of the effective potential.} and $\langle R\tilde{R}\rangle_\mathrm{ren}$ 
is the renormalized expectation value of the Chern-Pontryagin density $R{\tilde R}$, which is equivalent to the treatment
adopted in Ref.~\cite{Fischler:2007tj}. 
This is achieved by adding counter terms in the effective action ${\cal L}_\mathrm{ct}$ so that they cancel the divergences in $\langle R {\tilde R} \rangle$ as
$({\cal N}/64 \pi^2 M_\mathrm{Pl})\langle R {\tilde R} \rangle_\mathrm{ren} = ({\cal N}/64 \pi^2 M_\mathrm{Pl})\langle R {\tilde R} \rangle - (-g)^{-1/2} \delta ((-g)^{1/2} {\cal L}_\mathrm{ct})/\delta \phi$.  
By choosing the counter term 
\beq
\begin{split}
  {\cal L}_\mathrm{ct} = &\lmk a_1Rg^{\mu\nu}+ a_2R^{\mu\nu}\rmk M_{\rm Pl}^{4}\lmk\partial_{\mu}f\partial_{\nu}f\rmk\\
  &+\lmk a_3R^2 g^{\mu\nu}+ a_4 R R^{\mu\nu} + a_5R^{\rho\sigma}R_{\rho\sigma} g^{\mu\nu}\rmk M^{2}_{\rm Pl}  \lmk\partial_{\mu}f\partial_{\nu}f\rmk\\
  &+\lmk a_6R^3g^{\mu\nu} + a_7R(R^{\rho\sigma}R_{\rho\sigma})g^{\mu\nu} + a_8(R^{\rho\sigma}R_{\rho\sigma})R^{\mu\nu}\rmk\lmk\partial_{\mu}f\partial_{\nu}f\rmk,
\end{split}\label{counter1}
\eeq
where $a_i$ is dimensionless coefficient, and then 
substituting the background values of the metric and scalar fields, the relevant part of the effective equation of motion reads 
\beq
 \frac{1}{\sqrt{-g}}\frac{\delta (\sqrt{-g} {\cal L}_\mathrm{ct})}{\delta\phi}= \frac{{\mathcal N} H^2\tk}{64\pi^2M_{\rm Pl}^3}\lmk\alpha_1\Lambda^4+\alpha_2\Lambda^2H^2+\alpha_3\log\lmk2\Lambda/M_{\rm Pl}\rmk H^4 \rmk, \label{eomct}
\eeq
where $\alpha_i$ are linear combinations of $a_i$ with
\begin{equation}
a_1+\frac{a_2}{4} \sim \left(\frac{\Lambda}{M_{\rm Pl}}\right)^4 \alpha_1, \quad a_3+\frac{a_4}{4}+  \frac{a_5}{4}\sim  \left(\frac{\Lambda}{M_{\rm Pl}}\right)^2 \alpha_2, \quad a_6+\frac{a_7}{4}+\frac{a_8}{16} \sim \log    \left(\frac{2 \Lambda}{M_{\rm Pl}}\right) \alpha_3.
\end{equation}
Here we have assumed that  $\dot{\phi}$ is nearly constant. 
We can see that the terms in Eq.~\eqref{eomct} can cancel the UV cutoff dependent terms in  $\langle R {\tilde R} \rangle$ (Eq.~\eqref{RRlambda} or Eq.~\eqref{RRad2}) 
with the appropriate choice of the parameters $a_i$ so that we can absorb the UV divergences that appear in the right-hand side of Eq.~\eqref{efeom}. 
In this sense, we obtain the renormalized $R\tilde{R}$, which is necessary to make our assumption on the background dynamics consistent. 
We note that, on the background we have assumed, some terms degenerate with others, and therefore, we have ambiguity in choosing $a_i$. 
If we consider the renormalization in general background, the degeneracy would be resolved, but this is beyond the scope of this paper. 

\subsection{Implication to gravitational leptogenesis}
As we have seen in Sec.~\ref{leptogenesis}, lepton asymmetry can be produced during inflation 
when the chiral gravitational waves are generated
through the gravitational chiral anomaly~(Eq.~\eqref{G_anomaly}). We should stress that it is not yet clear if the value of renormalized $\langle R\tilde{R}\rangle$ can really be the right hand side of Eq. \eqref{G_anomaly}, since the renormalization of $\langle R\tilde{R}\rangle$ is done in the point of view of the effective
equation of motion for the $\phi$ field, which is independent of the lepton current that appears on the left hand side of Eq.~\eqref{G_anomaly}. Nevertheless we expect that the right hand side of the anomaly equation~\eqref{G_anomaly} can be properly renormalized in the same way as the $R\tilde{R}$ in the equation of motion of $\phi$~\eqref{efeom} so that the anomaly equation holds at the level of the renormalized expectation values.

As the simplest estimation, we take the minimal subtraction\footnote{This choice is also not covariant, but even if we took any covariant renormalization process, it would not change the order of the values of physical quantities.} for the UV divergences appearing in the expectation value 
of the Chern-Pontryagin density. 
The finite part of $R\tilde{R}$ in the cutoff regularization performed in Sec.~\ref{ana} is read from Eq.~\eqref{RRlambda} as\footnote{After subtracting $\Lambda$ dependence from the UV divergent expectation value in \eqref{RRlambda}, we find a finite term proportional to $H^6M^{-2}_{\rm Pl}\log(H/M_{\rm pl})$. Actually, within the parameter region we will consider, such a finite term becomes $\mathcal{O}(10)$ larger than that shown in \eqref{mini}. For simplicity, however, we subtract such contribution, which is consistently achieved by introducing a finite counter term. As we will see, including such a contribution does not change our evaluation of lepton number significantly.}
\beq
\la R\tilde{R}\ra_{\rm ren} = \frac{3}{16\pi^2}\frac{H^2}{M_{\rm Pl}^2}\Theta H^4\lmk \frac{39}{2} - 10\gamma\rmk.\label{mini}
\eeq
By integrating the anomaly equation with respect to $\eta$ over the inflationary era, the generated lepton asymmetry can be evaluated as
\beq
n_L(\eta_f)_{\rm ren} = \frac{N_{\rm R-L}}{24(4\pi)^2}\la R\tilde{R}\ra_{\rm ren}\frac{1}{3H}\lmk 1 -\frac{a^3_f}{a^3_i}\rmk \simeq -\frac{1}{2048\pi^4}\frac{H^2}{M_{\rm Pl}^2}\Theta H^3\lmk \frac{39}{2} - 10\gamma\rmk. 
\eeq
If the reheating temperature is higher than the electroweak scale, the electroweak sphaleron~\cite{Kuzmin:1985mm} converts this lepton asymmetry into the baryon asymmetry as
$n_B = - 28/79 \ n_L$~\cite{Harvey:1990qw} so that 
the resultant baryon-to-entropy ratio reads 
\beq
\frac{n_B}{s} = \frac{315}{80896\pi^6}\epsilon_{\rm w.o}\lmk\frac{\pi^2}{90}\rmk^{\frac{1}{1+w}}g_*^{-\frac{w}{1+w}}\lmk\frac{T_{\rm reh}}{M_{\rm Pl}}\rmk^{\frac{1-3w}{1+w}}\lmk\frac{H}{M_{\rm Pl}}\rmk^{\frac{3+5w}{1+w}}\Theta\lmk \frac{39}{2} - 10\gamma\rmk,\label{baryon}
\eeq
where we have parameterized the effective equation of state after the end of reheating before the completion of reheating as $p = w\rho$. 

In our previous work~\cite{Kamada:2019ewe}, we have found that the model with k-inflation accompanied by gravitational reheating realizing $w = 1$ and $T_{\rm reh} \sim 10^7$GeV successfully explain observed baryon asymmetry. See Eq.~\eqref{baryon_1}. 
However, the situation significantly changes when we take the minimal subtraction scheme for the renormalization. 
Using Eq.~\eqref{baryon}, we can evaluate the baryon asymmetry for $w = 1$ as
\beq
\frac{n_B}{s} = 1.3\times10^{-16}\lmk\frac{\epsilon_{\rm w.o}}{1}\rmk\lmk\frac{g_*}{100}\rmk^{-\frac{1}{2}}\lmk\frac{H}{10^{13}{\rm GeV}}\rmk^4\lmk\frac{\Theta}{10^{-5}}\rmk\lmk \frac{T_{\rm reh}}{10^2{\rm GeV}}\rmk^{-1}.\label{w=1}
\eeq
Compared to Eq.~\eqref{baryon_1}, parametrically, contributions proportional to $\mu^4$ disappear and instead we have contributions proportional to $H^4$. 
Thus the resultant baryon asymmetry is more suppressed about $(H/\mu)^4 \sim 10^{-12}$ according to the reference values. 
This is because the contributions proportional to $\Lambda^4$  
in the expectation value of the Chern-Pontryagin density $R{\tilde R}$ is removed by renormalization 
and only contributions of proportional to $H^4$  remain. 
Note that reheating should complete before the electroweak symmetry breaking, $T\sim 10^2$ GeV,  in order to convert the lepton asymmetry into the baryon asymmetry by the sphaleron process.\footnote{Such relatively low reheating temperature leads to too much gravitationally produced gravitons that is constrained by the BBN. If we take the reheating temperature higher so that it passes the BBN constraint, the resultant baryon asymmetry is much more suppressed.} As a result, baryon asymmetry can generate only $\sim 10^{-16}$ at most, which is far from the observed value. 
This indicates that gravitational leptogenesis in $\phi R\tilde{R}$ model cannot be a working scenario if the renormalization of $R\tilde{R}$ with the minimal subtraction scheme is taken into account, which is consistent with the conclusion in Ref.~\cite{Fischler:2007tj}. 
However, we should keep in mind that the finite part of the expectation value of the Chern-Pontryagin density $R\tilde{R}$ after renormalization and hence lepton number density depends on the renormalization condition, and we cannot exclude the possibility that relatively large contribution up to $\mu^4$ remains 
in the finite part. 
Here we have merely evaluated a specific value and pointed out that the model fails to explain the observed baryon asymmetry for a specific renormalization condition. The consequence of our calculation has just revealed the potential failure of the gravitational leptogenesis.

\section{Summary and discussion}\label{discuss}
In this paper, we have performed renormalization of the gravitational Chern-Pontryagin density, which is coupled to the pseudo-scalar inflaton, in the context of the gravitational leptogenesis scenario. By using Mellin-Barnes representation, we have specified the UV divergences of the expectation value of the Chern-Pontryagin density $R\tilde{R}$ by analytic calculations. Although the cutoff divergence can be simply found by the adiabatic regularization as we performed in Sec.~\ref{ad_reg}, the finite part can be obtained only by the calculation we performed in Sec.~\ref{ana}. This difference originates from the difference of the vacuum states chosen in each case. We have also identified the counter terms, which can remove all the UV divergent terms and are consistent with the previous study~\cite{Fischler:2007tj}. Finally, we have performed the renormalization and discussed how it affects the gravitational leptogenesis scenario. As already pointed out in Ref.~\cite{Fischler:2007tj}, the generated lepton number density becomes much smaller than the original result~\cite{Alexander:2004us} due to the removal of the cutoff dependent part. If we take the minimal subtraction scheme for the renormalization of $R\tilde{R}$, the lepton asymmetry is determined by the finite part which we obtained in the Sec.~\ref{reg}. As a result, we found that observed baryon asymmetry cannot be explained in this scenario. However, this does not mean that gravitational leptogenesis does not work in general. Since the renormalized value of the lepton number depends on the renormalization condition, the finite part has an ambiguity of $\mathcal{O}(\Lambda)$ if we allow any finite terms below the cutoff scale. Such an ambiguity is rather general in quantum field theory in curved spacetime. Although this ambiguity makes models less predictive, our conclusion that $\phi R\tilde{R}$ model with minimal subtraction fails to explain the baryon asymmetry of the universe casts a doubt for the viability of the gravitational leptogenesis. 
Let us argue the potential ways to fix this ambiguity. 
The renormalization condition is often determined by observations. 
In this sense, assuming the gravitational leptogenesis is responsible for the present baryon asymmetry of the Universe, 
we can determine the renormalization condition if we will detect the chirality in the stochastic gravitational waves in future. 
From the theoretical point of view, if there is a first principle to determine the UV complete theory without ambiguity,  
the renormalization condition and the finite counter terms 
can also be determined by the matching conditions between the UV complete theory and the IR effective theory with which we deal.

Finally, we should emphasize our assumption made in Sec.~\ref{reg}. We have assumed that $R\tilde{R}$ in the anomaly equation~\eqref{G_anomaly} can be renormalized in the same way as $R\tilde{R}$ in equation of motion of scalar field~\eqref{eom}. However, the counter terms which we have introduced in Sec.~\ref{ren} couples to the background scalar field, and the relation between these terms and the background fermions is not clear. For this reason, one might wonder whether $R\tilde{R}$ in Eq.~\eqref{G_anomaly} can be renormalized in the similar way. In order to clarify this point, further analysis of this system is required. Namely, we need to extend the system including pseudo-scalar and fermions being dynamical. Within such a framework, one would be able to find the relation between the dynamics of the scalar field and the evolution of the lepton asymmetry in a more satisfactory manner. We leave it for our future work.

\vskip 1cm
\noindent
{\large\bf Acknowledgements}\\
We are grateful to A.~Papageorgiou and A.~A.~Starobinsky for useful discussions. 
The work of KK was supported by JSPS KAKENHI, Grant-in-Aid for Scientific Research JP19K03842 
and Grant-in-Aid  for Scientific Research on Innovative Areas 19H04610. 
JK is supported by JSPS KAKENHI, Grant-in-Aid for JSPS Fellows 20J21866 and research program of the Leading Graduate Course for Frontiers of Mathematical Sciences and Physics (FMSP).
YY is supported by JSPS KAKENHI, Grant-in-Aid for JSPS Fellows JP19J00494.

\appendix
\section{Contour integrals}\label{appendix}
In this appendix, we perform the integration in Eq.~\eqref{integrand} in the same way as done in Ref.~\cite{Ballardini:2019rqh}. Note that the following calculation technique was originally developed in Refs.~\cite{Frob:2014zka,Kobayashi:2014zza}, see also~\cite{Adshead:2018oaa, Adshead:2019aac}. Here, we take the IR regulator $c \to 0$ since there is no IR divergence in our case as one can easily check. Then we define the following quantities,
\beq
I_1 \equiv \int_{\mathcal{C}_s}\int_{\mathcal{C}_t}\frac{ds}{(2\pi i)}\frac{dt}{(2\pi i)}f(s,t,\tilde{\kappa})\alpha^{s+t}\left\{\left(e^{-\pi\tilde{\kappa}-i\frac{\pi}{2}(s - t)} - e^{\pi\tilde{\kappa}+i\frac{\pi}{2}(s - t)}\right)\frac{(s+1)(t+1)}{s+t+3}\left(\frac{\Lambda}{H}\right)^{s+t+3}\right\},
\eeq
\beq
I_2 \equiv \int_{\mathcal{C}_s}\int_{\mathcal{C}_t}\frac{ds}{(2\pi i)}\frac{dt}{(2\pi i)}f(s,t,\tilde{\kappa})\alpha^{s+t}\left\{\left(e^{-\pi\tilde{\kappa}-i\frac{\pi}{2}(s - t)} + e^{\pi\tilde{\kappa}+i\frac{\pi}{2}(s - t)}\right)(-i)\frac{s-t}{s+t+4}\left(\frac{\Lambda}{H}\right)^{s+t+4}\right\},
\eeq
\beq
I_3 \equiv \int_{\mathcal{C}_s}\int_{\mathcal{C}_t}\frac{ds}{(2\pi i)}\frac{dt}{(2\pi i)}f(s,t,\tilde{\kappa})\alpha^{s+t}\left\{\left(e^{-\pi\tilde{\kappa}-i\frac{\pi}{2}(s - t)} + e^{\pi\tilde{\kappa}+i\frac{\pi}{2}(s - t)}\right)\frac{2st+3(s+t)+4}{s+t+4}\left(\frac{\Lambda}{H}\right)^{s+t+4}\right\},
\eeq
\beq
I_4 \equiv \int_{\mathcal{C}_s}\int_{\mathcal{C}_t}\frac{ds}{(2\pi i)}\frac{dt}{(2\pi i)}f(s,t,\tilde{\kappa})\alpha^{s+t}\left\{\left(e^{-\pi\tilde{\kappa}-i\frac{\pi}{2}(s - t)} - e^{\pi\tilde{\kappa}+i\frac{\pi}{2}(s - t)}\right)i\frac{2(s-t)}{s+t+5}\left(\frac{\Lambda}{H}\right)^{s+t+5}\right\},
\eeq
The expectation value of $R\tilde{R}$ can be written as
\beq
\langle R\tilde{R}\rangle = 3H^4\frac{\sinh^2(\pi\tilde{\kappa})}{\pi^4}\frac{H^2}{M^2_{\rm Pl}}(I_1 + I_2 + \frac{\Theta}{16}(-I_3 + I_4)).
\eeq
Note that they can be written in the following form,
\beq
\begin{split}
  I_{1,2}(\kappa) &= \frac{1}{2}\left(I_{1,2}(\kappa) - I_{1,2}(-\kappa)\right)  \ \left(= \mathcal{O}(\kappa^{-1}) + \mathcal{O}(\kappa) + \mathcal{O}(\kappa^3)\right), \\
  I_{3,4}(\kappa) &= \frac{1}{2}\left(I_{3,4}(\kappa) + I_{3,4}(-\kappa)\right) \ \left(= \mathcal{O}(\kappa^{-2}) + \mathcal{O}(\kappa^0) + \mathcal{O}(\kappa^2)\right). 
\end{split}
\eeq

\begin{figure}[H]
  \centering
  \includegraphics[width=5cm,clip]{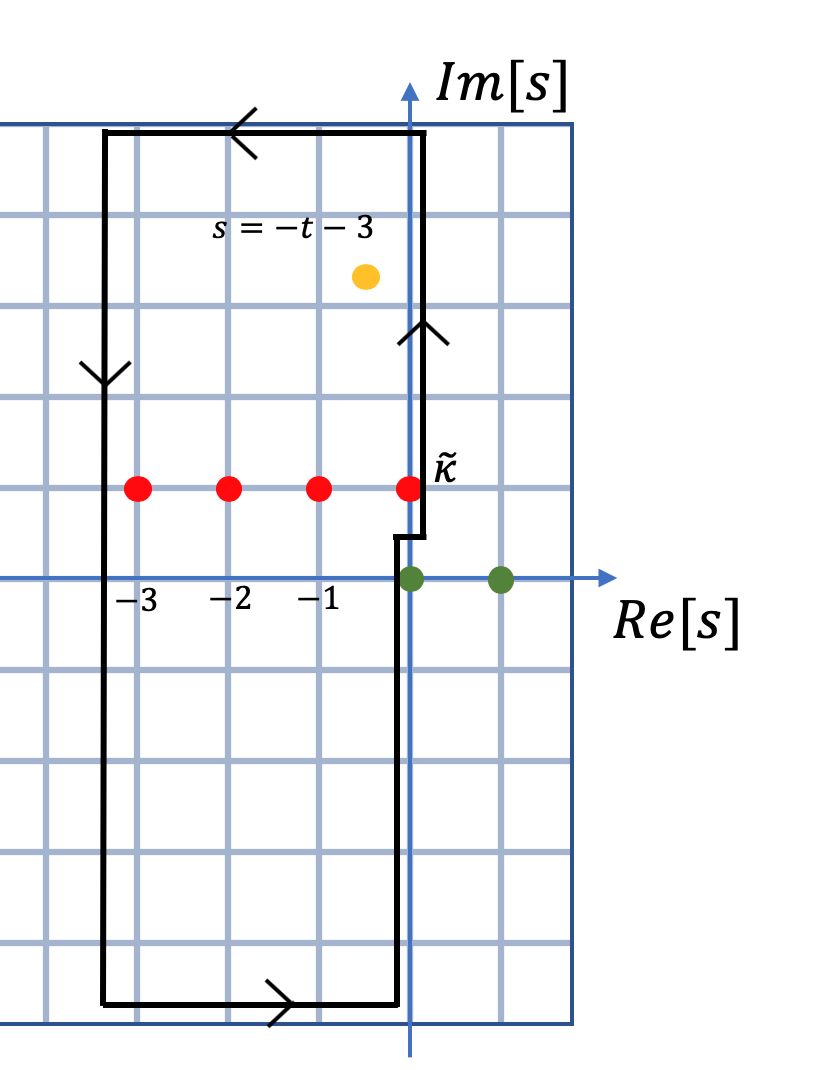}
    \caption{Integration contour $\mathcal{C}_ s$ for $I_1$.}
    \label{contour1}
\end{figure}

Let us explicitly calculate $I_1$. At first, we perform integration with respect to $s$.
Here contour $\mathcal{C}_s$ is defined to run from $-i\infty$ to $+i\infty$ and is chosen to separate the poles of $\Gamma(-s-1)$ and $\Gamma(-s+2)$ from those of $\Gamma(s - i\tilde{\kappa})$.
Since $(\Lambda/H)^{(s+t+n)}$ goes to zero for ${\rm{Re}}[s+t+n] < 0$ with $\Lambda \to \infty$, it is better to choose a closed contour in the left half-plane for both variables, $s$ and $t$. Assuming that ${\rm Re}[s] \lesssim 0$ and ${\rm Re}[t] \lesssim 0$,  $(\Lambda/H)^{(s+t+3)}$ vanishes for ${\rm Re}[s] < -3$. Thus, we consider the contour shown in Fig.~\ref{contour1}. Note that the integrand vanishes for $|{\rm Im}[s]|\to\infty$ due to the asymptotic property of Gamma-function.

As a result, the integral is $2\pi i$ times the sum of the residues of the following poles:
\beq
s = i\tilde{\kappa}, i\tilde{\kappa}-1, i\tilde{\kappa}-2, i\tilde{\kappa}-3, -t-3.
\eeq
Then, $I_1$ can be decomposed as
\beq
I_1 =  I_{1,\Lambda} + I_{1, s = -t-3},
\eeq
where
\beq
\begin{split}
  I_{1, \Lambda} \equiv \frac{1}{2}\sum_{n = 0}^3&\int_{\mathcal{C}_t}\frac{dt}{2\pi i}\alpha^{t+i\tilde{\kappa}-n}\frac{(-1)^n}{n!}\Gamma(n-1-i\tilde{\kappa})\Gamma(n+2-i\tilde{\kappa})\Gamma(t+i\tk)\Gamma(-t-1)\Gamma(-t+2)\\
  &\times\left(e^{\frac{\pi}{2}(it-\tk+ni)}-e^{-\frac{\pi}{2}(it-\tk+ni)}\right)\frac{(t+1)(i\tk-(n-1))}{t-(n-3-i\tk)}\left(\frac{\Lambda}{H}\right)^{t-n+3+i\tk}\\
  &-(\kappa\to-\kappa)\label{1lambda}
\end{split}
\eeq
and
\beq
\begin{split}
  I_{1, s = -t-3} \equiv \frac{1}{2}&\int_{\mathcal{C}_t}\frac{dt}{2\pi i}\alpha^{-3}\Gamma(-t-3-i\tilde{\kappa})\Gamma(t+i\tk)\Gamma(t+2)\Gamma(t+5)\Gamma(-t-1)\Gamma(-t+2)\\
  &\times(-i)\left(e^{\pi i(t+i\tk)}+e^{-\pi i(t+i\tk)}\right)(-t-2)(-t+1)\\
  &-(\kappa\to-\kappa)\\
  =  \frac{1}{2}&\int_{\mathcal{C}_t}\frac{dt}{2\pi i}\alpha^{-3}(-i)\frac{\pi^3\left(e^{\pi i(t+i\tk)}+e^{-\pi i(t+i\tk)}\right)}{\sin(\pi(t+i\tk))\sin^2\pi t}\frac{(t+4)(t+3)(t+2)^2(t+1)^2t(t-1)}{(t+i\tk)(t+i\tk+1)(t+i\tk+2)(t+i\tk+3)}\\
&-(\kappa\to-\kappa)\label{s+t+3=0}
\end{split}
\eeq
We then evaluate Eq.~\eqref{1lambda}. The integration with respect to $t$ is performed in the same way as above. As an example, let us calculate the $n = 0$ term,
\beq
\frac{i}{\alpha^3}\left[\int_{\mathcal{C}_t}\frac{dt}{2\pi i}\Gamma(-i\tk)\Gamma(2-i\tk)\Gamma(-t)\Gamma(-t+2)\sin(\frac{\pi}{2}(t+i\tk))\frac{\Gamma(t+i\tk)}{t+i\tk+3}\lmk\frac{\alpha\Lambda}{H}\rmk^{t+i\tk+3} -(\tk\to -\tk)\right].
\eeq
The relevant poles are $t = -1-i\tk, -3-i\tk$. Contribution from $t=-1-i\tk$ is
\beq
\frac{i}{16}\lnk\Gamma(1+i\tk)\Gamma(3+i\tk)\Gamma(-i\tk)\Gamma(2-i\tk) -(\tk\to-\tk)\rnk\lmk\frac{2\Lambda}{H}\rmk^2 = -\frac{\tk(1+\tk^2)\pi^2}{\sinh^2[\pi\tk]}\lmk\frac{\Lambda}{H}\rmk^2.
\eeq
The other contribution can be found as
\beq
\begin{split}
  &\frac{i}{48}\lnk \Gamma(3+i\tk)\Gamma(5+i\tk)\Gamma(-i\tk)\Gamma(2-i\tk)\lmk \psi(3+i\tk)+\psi(5+i\tk)+\gamma-\log\lmk\frac{2\Lambda}{H}\rmk-\frac{11}{6}\rmk- (\tk\to-\tk)\rnk \\
  &=\frac{-1}{48\tk}\lmk344-96\gamma-96\log\lmk\frac{2\Lambda}{H}\rmk\rmk,
  \end{split}
\eeq
where $\gamma$ is Euler's constant and $\psi(x)$ is the digamma function.

\begin{figure}[H]
  \centering
  \includegraphics[width=5cm,clip]{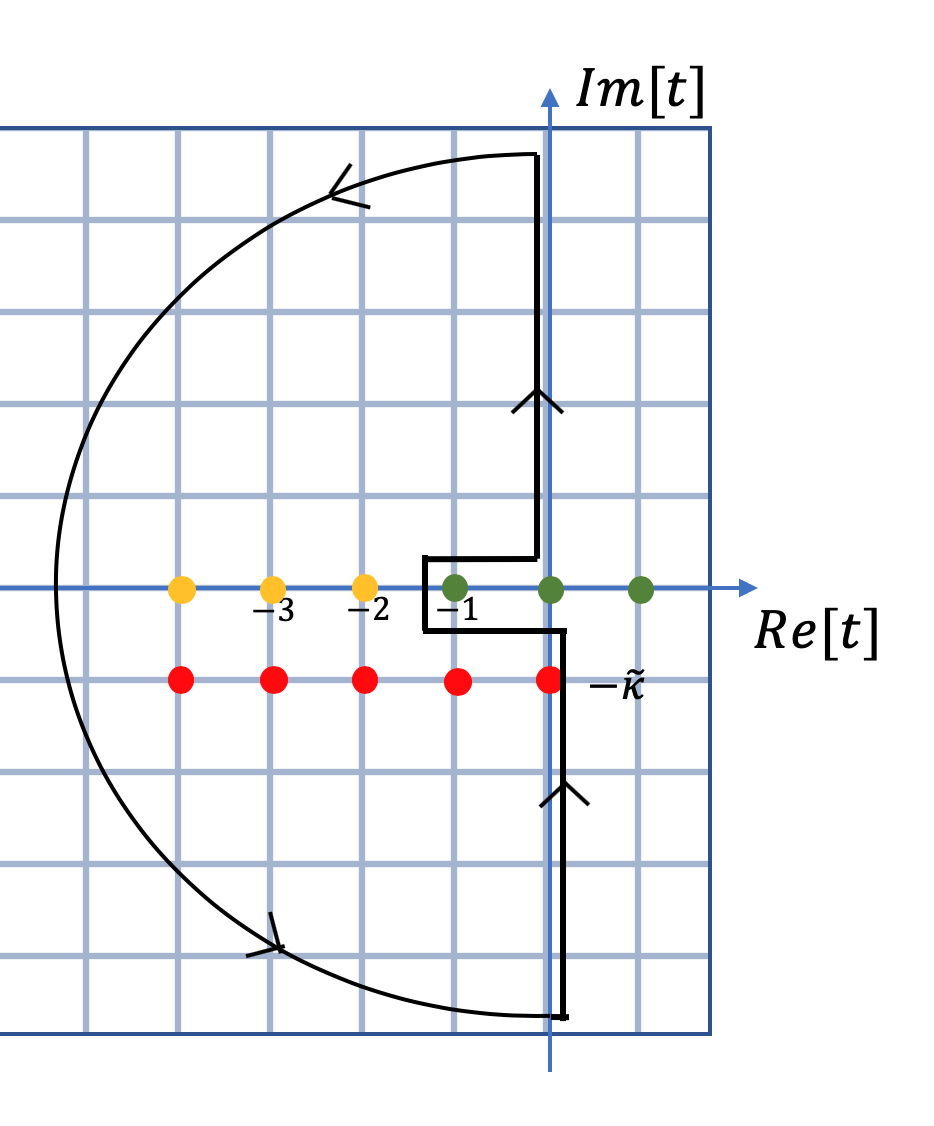}
    \caption{Integration contour $\mathcal{C}_t$ for terms proportional to $a_r/(t+i\tk)$.}
    \label{contour3}
\end{figure}

\begin{figure}[H]
  \centering
  \includegraphics[width=5cm,clip]{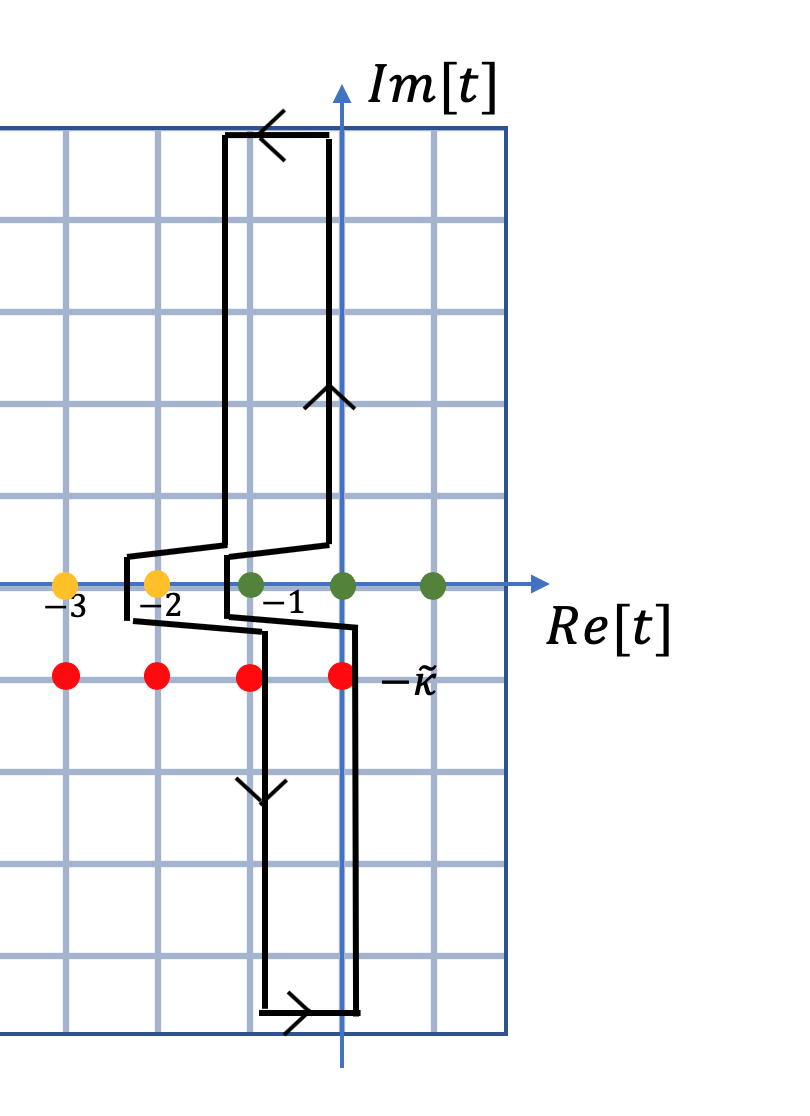}
    \caption{Integration contour $\mathcal{C}_t$ for remaining terms.}
    \label{contour2}
\end{figure}

Next, we evaluate Eq.~\eqref{s+t+3=0}. We decompose $I_{1, s=-t-3}$ into the following form,
\beq
I_{1, s=-t-3} = \frac{-i\pi^3}{8}\int_{\mathcal{C}_t}\frac{dt}{2\pi i}\frac{\cos(\pi i(t+i\tk))}{\sin(\pi(t+i\tk))\sin^2\pi t}\left\{ \frac{a_r}{t+i\tk}+B_r(t) - B_r(t-1)\right\} -(\tk \to  -\tk),\label{s=-t-3}
\eeq
where $B_r(s)$ is defined as
\beq
B_r(t) = \frac{b_{r,1}}{t+i\tk+1}+\frac{b_{r,2}}{t+i\tk+2}+\frac{b_{r,3}}{t+i\tk+3}+b_{r,4}t + b_{r,5}t^2 + b_{r,6}t^3 + b_{r,7}t^4 + b_{r,8}t^5
\eeq
and the coefficients $a_r, b_{r,n}$ are independent of $t$. The integration of Eq.~\eqref{s=-t-3} is performed along the contours shown in Fig.~\ref{contour3} and Fig.\ref{contour2}. Note that for a term proportional to $a_r$, we redefine the integrand as
\beq
\lim_{p\to 1}\frac{\cos(\pi(t+i\tk))}{\sin(\pi(t+i\tk))\sin^2\pi t}\frac{a_r}{(t+i\tk)^p}.
\eeq


The contribution from the integration contour shown in Fig.~\ref{contour3} is
\beq
\begin{split}
  &\frac{-i}{8}\frac{-\pi^2}{\sinh^2(\pi\tk)}\left\{-a_r(\tk)(\psi(2-i\tk)+\gamma)+a_r(-\tk)(\psi(2+i\tk)+\gamma)\right.\\
  &\left.-\frac{i}{2\pi}a_r(\tk)\sinh(2\pi\tk)\psi^{(1)}(2-i\tk)-\frac{i}{2\pi}a_r(-\tk)\sinh(2\pi\tk)\psi^{(1)}(2+i\tk)\right\}\\
  &=\frac{\pi^2}{\sinh^2(\pi\tk)}4\tk+\mathcal{O}(\tk),
\end{split}
\eeq
with $a_r(\tk) = 8i(2\tk-5\tk^3-7\tk^5)$. The contribution from remaining terms is
\beq
\begin{split}
  &\frac{-i}{8}\frac{-\pi^2}{\sinh^2(\pi\tk)}\left\{\frac{4}{15}\left(2i\tk(1+\tk^2)(97 -307\tk^2-5\tk^4)-\frac{i(92-143\tk^2+95\tk^4-210\tk^6)}{(1+\tk^2)\pi}\sinh(2\pi\tk)\right)\right\}\\
  &=-\frac{1}{3\tk}+\mathcal{O}(\tk).
\end{split}
\eeq
By combining them all and picking leading order in terms of $\tk$, we finally obtain $I_1$ as
\beq
\frac{\sinh^2(\pi\tk)}{\pi^2}I_1 = \tk\lmk-2\lmk\frac{\Lambda}{H}\rmk^2+4\gamma+4\log\lmk\frac{2\Lambda}{H}\rmk-7\rmk+\mathcal{O}(\tk^2).
\eeq

The integration of $I_2, I_3$ and $I_4$ can be done in the same way. At the leading order in $\tk$, we obtain
\beq
\frac{\sinh^2(\pi\tk)}{\pi^2}I_2 = \tk\lmk\lmk\frac{\Lambda}{H}\rmk^4+5\lmk\frac{\Lambda}{H}\rmk^2-14\gamma-14\log\lmk\frac{2\Lambda}{H}\rmk+\frac{53}{2}\rmk+\mathcal{O}(\tk^2),
\eeq

\beq
\frac{\sinh^2(\pi\tk)}{\pi^2}\frac{\Theta}{16}I_3 = 2\tk\lmk\frac{\Lambda}{H}\rmk^4+\mathcal{O}(\tk^2),
\eeq

\beq
\frac{\sinh^2(\pi\tk)}{\pi^2}\frac{\Theta}{16}I_4 = 2\tk\lmk\frac{\Lambda}{H}\rmk^4 + \mathcal{O}(\tk^2).
\eeq

As a result, the expectation value of $R\tilde{R}$ is found to be
\beq
\begin{split}
  \langle R\tilde{R}\rangle &= 3H^4\frac{\sinh^2(\pi\tilde{\kappa})}{\pi^4}\frac{H^2}{M^2_{\rm Pl}}\left(I_1 + I_2 + \frac{\Theta}{16}(-I_3 + I_4)\right) \\
  &= 3H^4\frac{H^2}{\pi^2M_{\rm Pl}^2}\tk\lnk\lmk\frac{\Lambda}{H}\rmk^4+3\lmk\frac{\Lambda}{H}\rmk^2-10\log\lmk\frac{2\Lambda}{H}\rmk-10\gamma+\frac{39}{2}\rnk+\mathcal{O}(\tk^2).
\end{split}
\eeq

\bibliographystyle{JHEP}
\bibliography{RRR_ref}
\end{document}